\documentclass[12pt,epsf]{article}
\usepackage{graphicx,amsmath,amssymb,booktabs,enumerate}
\usepackage{subfigure}
\usepackage{comment}
\usepackage{color}

\begin{document}

\def\a{\alpha}
\def\b{\beta}
\def\c{\varepsilon}
\def\d{\delta}
\def\e{\epsilon}
\def\f{\phi}
\def\g{\gamma}
\def\h{\theta}
\def\k{\kappa}
\def\l{\lambda}
\def\m{\mu}
\def\n{\nu}
\def\p{\psi}
\def\q{\partial}
\def\r{\rho}
\def\s{\sigma}
\def\t{\tau}
\def\u{\upsilon}
\def\v{\varphi}
\def\w{\omega}
\def\x{\xi}
\def\y{\eta}
\def\z{\zeta}
\def\D{\Delta}
\def\G{\Gamma}
\def\H{\Theta}
\def\L{\Lambda}
\def\F{\Phi}
\def\P{\Psi}
\def\S{\Sigma}

\def\o{\over}
\def\beq{\begin{eqnarray}}
\def\eeq{\end{eqnarray}}
\newcommand{\gsim}{ \mathop{}_{\textstyle \sim}^{\textstyle >} }
\newcommand{\lsim}{ \mathop{}_{\textstyle \sim}^{\textstyle <} }
\newcommand{\vev}[1]{ \left\langle {#1} \right\rangle }
\newcommand{\bra}[1]{ \langle {#1} | }
\newcommand{\ket}[1]{ | {#1} \rangle }
\newcommand{\EV}{ {\rm eV} }
\newcommand{\KEV}{ {\rm keV} }
\newcommand{\MEV}{ {\rm MeV} }
\newcommand{\GEV}{ {\rm GeV} }
\newcommand{\TEV}{ {\rm TeV} }
\def\diag{\mathop{\rm diag}\nolimits}
\def\Spin{\mathop{\rm Spin}}
\def\SO{\mathop{\rm SO}}
\def\O{\mathop{\rm O}}
\def\SU{\mathop{\rm SU}}
\def\U{\mathop{\rm U}}
\def\Sp{\mathop{\rm Sp}}
\def\SL{\mathop{\rm SL}}
\def\tr{\mathop{\rm tr}}

\def\IJMP{Int.~J.~Mod.~Phys. }
\def\MPL{Mod.~Phys.~Lett. }
\def\NP{Nucl.~Phys. }
\def\PL{Phys.~Lett. }
\def\PR{Phys.~Rev. }
\def\PRL{Phys.~Rev.~Lett. }
\def\PTP{Prog.~Theor.~Phys. }
\def\ZP{Z.~Phys. }


\baselineskip 0.7cm

\begin{titlepage}

\begin{flushright}
IPMU-12-0134\\
UT-12-19
\end{flushright}

\vskip 1.35cm
\begin{center}
{\large \bf Gluino Decay as a Probe of \\
High Scale Supersymmetry Breaking
}
\vskip 1.2cm
Ryosuke Sato${}^{1,2}$, Satoshi Shirai${}^{3,4}$ and Kohsaku Tobioka${}^{1,2}$
\vskip 0.4cm

{\it

$^1$Department of Physics, University of Tokyo,\\
Tokyo 113-0033, Japan,\\
$^2$Kavli Institute for the Physics and Mathematics of the
  Universe (WPI), \\Todai Institutes for Advanced Study, University of Tokyo,\\
  Kashiwa 277-8583, Japan\\
$^3$Department of Physics, University of California, \\Berkeley, CA 94720, USA\\
$^4$ Theoretical Physics Group, Lawrence Berkeley National Laboratory, \\Berkeley, CA 94720, USA

}

\vskip 1.5cm

\abstract{
A supersymmetric standard model with heavier scalar supersymmetric particles has many attractive features.
If the scalar mass scale is ${\cal O}(10-10^4)$ TeV, the standard model like Higgs boson with mass around 125 GeV,
 which is strongly favored by the LHC experiment, can be realized.
However, in this scenario the scalar particles are too heavy to be produced at the LHC.
In addition, if the scalar mass is much less than ${\cal O}(10^4)$ TeV, the lifetime of the gluino is too short to be {measured}.
Therefore, it is hard to probe the scalar particles at a collider.
However, a detailed study of the gluino decay reveals that two body decay of the gluino carries important information on the
scalar scale.
In this paper, we propose a test of this scenario by measuring the decay pattern of the gluino at the LHC.
}
\end{center}
\end{titlepage}

\setcounter{page}{2}


\section{Introduction}
A supersymmetric (SUSY) standard model (SSM) is the most promising model beyond the standard model (SM).
At the Large Hadron Collider (LHC) experiment,  the Higgs boson search of the ATLAS and CMS collaborations reveals the existence of the Higgs boson-like particle with mass around 125 GeV \cite{ATLAS,CMS, 20120704}.
However, the minimal SSM (MSSM) naturally predicts relatively light mass of the Higgs boson, and thus
the parameter region where the Higgs mass is heavy is strongly constrained. 
To realize the Higgs mass about 125 GeV, a heavy scalar top quark (stop) and/or large $A$-term of the stop are required,
in the case of the MSSM \cite{Higgs,Heinemeyer:2011aa,Arbey:2011ab}.

Among them, heavy scalar scenarios are interesting from many viewpoints.
The Higgs mass around 125 GeV indicates the supersymmetric scalar mass scale of ${\cal O}(10-10^4)$ TeV \cite{Arbey:2011ab,Giudice:2011cg,Ibe:2011aa,Ibe:2012hu}.
Such a heavy scalar particles can greatly relax the notorious constraints of the SUSY flavor 
and CP problem and the cosmological gravitino problem
as discussed in the context of the split SUSY \cite{ArkaniHamed:2004fb,Giudice:2004tc,ArkaniHamed:2004yi},
the anomaly mediated SUSY breaking (AMSB) models \cite{Randall:1998uk, Giudice:1998xp} and
the spread SUSY \cite{Hall:2011jd}.
From a theoretical point of view, it is more likely that the scalars have larger mass compared to the fermions, 
such as gauginos and Higgsinos. 
This is because the mass of fermions can be protected from a large radiative corrections thanks to some
symmetry and, on the other hand, that of the scalar particle is not.
Therefore the MSSM with heavy scalar particles is favored from both phenomenological and theoretical viewpoints.

In order to test this scenario, an estimation of the scalar mass scale, especially for the stop mass, is quite important, since
it strongly connects to the Higgs mass.
This also provides insights into the low-energy precision measurements, such as flavor and CP physics.
However, the scalar mass is too heavy to be produced at the colliders in this scenario, even if
the gauginos are light enough to be produced.
Therefore we need indirect information to study the scalar sector. 
For example such heavy particles can be relevant in the early universe, and,
in some case, this mass scale can be determined by a gravitational wave \cite{Saito:2012bb}.
A precise measurement of gaugino-Higgsino mixings at the International Linear Collider (ILC) may enable us to
estimate the scalar mass scale by using renormalization group (RG) equations \cite{Kilian:2004uj}.
At the LHC, if the scalar mass scale is much larger than about $10^{3-4}$ TeV,
an R-hadron or a displaced vertex of a long-lived gluino is expected \cite{ArkaniHamed:2004fb,Kilian:2004uj,Toharia:2005gm}.
The lifetime of the gluino is very sensitive to the scalar mass scale, and hence
measurement of the lifetime of the gluino will be a probe of such a heavy scalar mass. 
However, this method cannot be applicable in the case that the scalar mass scale is much smaller than ${\cal O}(10^{4})$ TeV,
since the decay length is too short to be {measured} at a collider experiment.

In this paper, we study a possibility to test the scalar mass scale in the heavy scalar scenarios 
through the gluino decay patterns.
In the case of the short-lived gluino, what we can measure is basically branching fractions of the gluino.
If we study only the gluino three body decay,\footnote{Previously, the three body decay of the gluino in the heavy scalar scenario is studied in  the different context. See, for examples, Refs.\cite{threebody}.
} 
it is difficult to extract the information of the scalar mass scale.  
On the other hand, the two body decay of the gluino is sensitive to the scalar mass scale, as we will see in the next section.
Actually, the Higgs mass around 125 GeV implies a possibility of a large size of two body decay of the gluino. 
We discuss that the branching fraction of $ {\tilde g} \to g \tilde{\chi}^0$ carries important information of
the scalar mass scale, and we propose a simple method to measure it. This method relies on a naive counting of the events which are classified with their 
jet, lepton and $b$-jet multiplicities and kinematical behaviors of jets and missing energy. 
In order to define signal regions, we only use commonly-used cuts adopted in the study of the ATLAS and CMS collaborations. 
We show that the branching fraction can be determined to accuracy of 10 \% 
if the ${\cal O}(10^4)$ gluinos are produced at the LHC.

This paper is organized as follows. We first review the gluino decay in the case of the heavy scalars in Section 2. 
We propose a basic strategy to discriminate the two body decay of the gluino and demonstrate it for several models in Section 3. 
The conclusion and discussion are given in Section 4. 


\section{Heavy scalar scenario and gluino decay}
In this section, we briefly review the gluino decays in the heavy scalar scenario.
First, we see the decay width of the two body decay of the gluino has
a logarithmic enhancement factor compared to the three body decay
by using leading order calculation \cite{Toharia:2005gm, gluinodecay}.
Thanks to this behavior, we can probe the squark mass scale with the branching fraction of the gluino decay.
However, when the squarks are much heavier than the gluino,
it is necessary to resum leading logarithmic corrections for the precise determination of the decay width of the gluino \cite{Gambino:2005eh}.
Therefore, next, we discuss the RG-improved calculation of the decay width according to Ref. \cite{Gambino:2005eh}.
In the following of this paper, we neglect the left-right mixings of the squarks,
because the mixing angles are suppressed with quark masses and are negligible in the heavy scalar scenario.

\subsection{Leading order calculation}

\subsubsection*{Three body decay ($\tilde g \to q\bar q \tilde\chi$)}

\begin{figure}[t!]
\begin{center}
\subfigure[Three body decay]{
\includegraphics[scale=0.605]{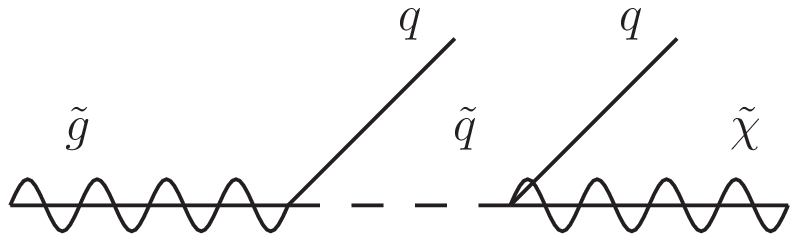}}
\subfigure[Two body decay]{
\includegraphics[scale=0.605]{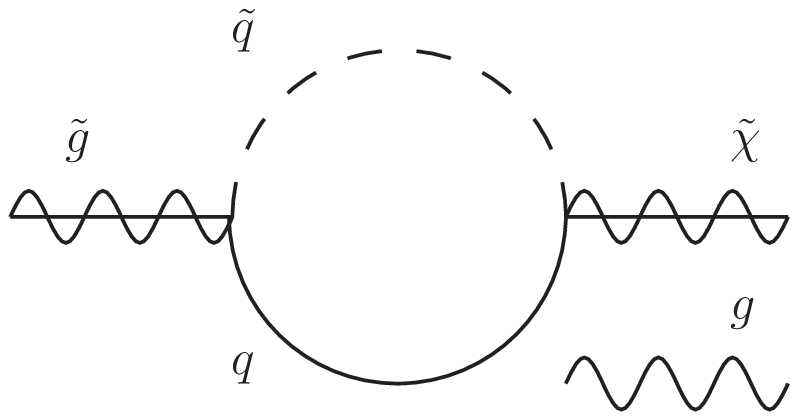}
}
\caption{
Examples of diagrams relevant for the gluino decay.
}\label{fig:decay-d}

\end{center}
\end{figure}
Squark exchanging diagrams at tree level induce three body decays of the gluino (Fig. \ref{fig:decay-d}-(a)).
We assume the squark sector respects flavor symmetries for simplicity.\footnote{
In the heavy scalar scenario, the flavor violation in the squark sector is less constrained,
because the dangerous flavor-changing-neutral-current (FCNC) processes are suppressed by the large squark masses.
However, in spite of the heavy squark,
the gluino three body decay can be a good probe of the flavor violation in the squark sector,
because three body decay is a direct result of the four-Fermi interaction.
We will study this possibility in more detail elsewhere \cite{fv_in_gluino}.
}
For massless quark limit, we can get an analytical expression of the gluino partial decay width {\cite{Gambino:2005eh, 3body_formula}}, e.g., 
\begin{eqnarray}
\G(\tilde g\to q_L q_R^c \tilde B) &=& \frac{g_s^2 g'^2 Y_{q_L}^2 }{{1536}\pi^3} \frac{m_{\tilde g}^5}{m_{{\tilde q}_L}^4}
\left[ 
 f\left( \frac{m_{\tilde B}^2}{m_{\tilde g}^2}\right)
+ \frac{2m_{\tilde B}}{m_{\tilde g}} ~g\left( \frac{m_{\tilde B}^2}{m_{\tilde g}^2}\right)
 \right],
\end{eqnarray}
where 
$f(x) = 1 -8x -12x^2 \log x +8x^3 -x^4$ and
$g(x) = 1 + 9x + 6x\log x - 9x^2 + 6x^2\log x- x^3$.
As we can see from the above expression,
the decay widths simply damp as $m_{\tilde q}^{-4}$ in the heavy squark limit.

\subsubsection*{Two body decay ($\tilde g \to g \tilde\chi^0$)}
The gluino can decay into a neutralino and a gluon via quark-squark loop diagrams (Fig. \ref{fig:decay-d}-(b)).
First, one can easily find a branching fraction to a Wino ($\tilde W$) is strongly suppressed,
because such a decay mode is induced by dimension 7 effective interactions, such as,
\begin{eqnarray}
{\cal L}_{\rm eff.} \simeq \frac{m}{m_{\tilde q_L}^4} (H^\dagger \t^a H) \tilde W^a \s^{\m\n} \tilde g G_{\m\n}.
\end{eqnarray}
Then, $\tilde g \to g \tilde W$ is mainly induced by the mixing with other electroweak(EW)-inos.

As for the gluino decay into a Bino ($\tilde B$) or a Higgsino ($\tilde h$),
the decay rates
$\G(\tilde g\to g\tilde B)$ and $\G(\tilde g\to g\tilde h)$ are given by,
\begin{eqnarray}
\G(\tilde g \to g\tilde B) &\simeq& 
\frac{g'^2 g_s^4}{32786\pi^5}
\frac{(m_{\tilde g}^2-m_{\tilde B}^2)^3}{m_{\tilde g}^3} 
\left( \sum_q \frac{Y_{q_L}}{m_{\tilde q_L}^2} - \frac{Y_{q_R}}{m_{\tilde q_R}^2} \right)^2 (m_{\tilde g}-m_{\tilde B})^2 ,\label{eq:to_bino} \\
\G(\tilde g \to g\tilde h) &\simeq& 
\frac{\hat y_t^2 g_s^4}{4096\pi^5}
\frac{(m_{\tilde g}^2-m_{\tilde h}^2)^3}{m_{\tilde g}^3}
 \Biggl[
 \frac{m_t}{m_{\tilde t_L}^2} \left( \log\frac{m_{\tilde t_L}^2}{m_t^2} -1\right)
 + \frac{m_t}{m_{\tilde t_R}^2} \left( \log\frac{m_{\tilde t_R}^2}{m_t^2} -1\right)
  \Biggr]^2. \label{eq:to_higgsino}
%
\end{eqnarray}
Here, $\hat y_t$ is Yukawa coupling constant in the MSSM.
The index $q$ in Eq. (\ref{eq:to_bino}) runs through $u,d,s,c,b$ and $t$, and
$Y_q$ is the hypercharge of the quarks.
In Eq. (\ref{eq:to_higgsino}), we neglect contributions other than top-stop loops,
because it has a dominant contribution.
Eq. (\ref{eq:to_higgsino}) shows
$\G(\tilde g \to g\tilde h)$ has an enhancement factor $m_t^2 / m_{\tilde g}^2 ( \log(m_{\tilde t}^2 / m_t^2) )^2$,
although $\G(\tilde g\to g\tilde B)$ and $\G(\tilde g \to q\bar q \tilde\chi)$ simply damp as $m_{\tilde q}^{-4}$. %
Then, a rough behavior of the ratio of $\G(\tilde g \to g\tilde h)$ to $\G(\tilde g \to q\bar q\tilde\chi)$ is given by,
\begin{eqnarray}
\frac{ \G(\tilde g\to g\tilde h) }{ \G(\tilde g \to q\bar q \tilde \chi) }
\propto
\frac{m_t^2}{m_{\tilde g}^2} \left( \log\frac{m_{\tilde t}^2}{m_t^2} \right)^2.\label{eq:B_ratio}
\end{eqnarray}
Therefore, we can expect the branching fraction of the two body decay of the gluino is
significantly enhanced in the heavy scalar scenario.
However, in this scenario,
it is important to resum the leading logarithm corrections,
then, we have to use an RG-improved calculation.
{
Furthermore, RG-improved calculation is helpful to understand
the origin of the logarithmic correction in Eq. (\ref{eq:to_higgsino}).
As a similar example, in case that some particles much heavier than others,
the RG evolution of coupling constants generates sizable logarithmic correction as a hard SUSY breaking effect
\cite{Cheng:1997sq, Nojiri:1997ma, Katz:1998br}.
In the next subsection we discuss the RG calculation leads milder dependence on the scalar mass, compared to Eq. (\ref{eq:B_ratio}).
}

\subsection{RG-improved calculation}
In the present model, the squark mass scale is much larger than the gluino mass.
Thus,
as we denoted in the end of the previous subsection,
the above leading order calculation is not suitable
for the calculation of the gluino decay width.
To calculate them more precisely, we integrate out the squarks
and use the effective Lagrangian
which describes the dynamics of the SM particles together with Higgsinos and gauginos.
In this setup, the gluino decay is caused by non-renormalizable interactions which are induced by integrating out the squarks.
For example, a four Fermi-type operator induces the three body decay of the gluino,
and a dipole-type operator the two body decay.
To resum the logarithmic corrections of the decay width, 
RG equations for the Wilson coefficient are useful.
The RG equations which concern with the gluino decay in the heavy scalar scenario
are calculated by Gambino et. al. in Ref. \cite{Gambino:2005eh},
and we use them in the following of this paper.

RG equations are also useful to understand the logarithmic enhancement in Eq. (\ref{eq:to_higgsino})
from the view point of the effective theory.
Then, let us discuss the Wilson coefficients which concern with the two body decay.
They are given by,
\begin{eqnarray}
{\cal L}_{\rm eff.} = C_7^{\tilde B} Q_7^{\tilde B} + \left( C_2^{\tilde H} Q_2^{\tilde H} + C_5^{\tilde H} Q_5^{\tilde H} + {\rm h.c.} \right).
\end{eqnarray}
Here, $Q$'s are operators defined below:
\begin{eqnarray}
Q_7^{\tilde B} &=& {\bar {\tilde B}} \s^{\m\n} \g^5 {\tilde g} ~G_{\m\n},\\
Q_2^{\tilde H} &=& \left({\bar {\tilde H}} \s^{\m\n} P_R {\tilde g}^a \right) \left( \bar q^{(3)} \s_{\m\n}  T^a P_R t \right), \\
Q_5^{\tilde H} &=& {\bar {\tilde H}} \s^{\m\n} P_R {\tilde g} h ~G_{\m\n}.
\end{eqnarray}
%
We have to match the low energy effective theory to MSSM at some matching scale $\tilde m$,
which is set to a typical scale of the squark masses.
The Wilson coefficients at the matching scale $\tilde m$ are given by,
\begin{eqnarray}
C_7^{\tilde B}(\tilde m) &=& \frac{g_s^2 g'}{128\pi^2} (m_{\tilde g}-m_{\tilde B}) \left(\sum_q \frac{Y_{q_L}}{m_{\tilde q_L}^2} - \frac{Y_{q_R}}{m_{\tilde q_R}^2}\right), \label{eq:c7b}\\
C_2^{\tilde H}(\tilde m) &=& \frac{g_s y_t}{4\sqrt{2}\sin\b} \left(\frac{1}{m_{\tilde q_L^{(3)}}^2} + \frac{1}{m_{\tilde t_R}^2} \right), \\
C_5^{\tilde H}(\tilde m) &=& \frac{g_s^2 y_t^2}{32\sqrt{2}\pi^2\sin\b} \left(\frac{1}{m_{\tilde q_L^{(3)}}^2} + \frac{1}{m_{\tilde t_R}^2} \right).
\end{eqnarray}
Here, $q$ in Eq. (\ref{eq:c7b}) runs through $u, d, s, c, b$ and $t$.
$y_t$ is a coupling constant for top-top-Higgs coupling like the SM.
We can see that $C_7^{\tilde B}(\tilde m)$ and $C_5^{\tilde H}(\tilde m)$ are suppressed
by a loop factor compared to $C_2^{\tilde H}(\tilde m)$.
Below the scale $\tilde m$, the running of the Wilson coefficients is described by the following equations:
\begin{eqnarray}
16\pi^2 \m \frac{d}{d\m} C_7^{\tilde B} &=& -14 g_s^2 C_7^{\tilde B}, \label{eq:rge_cb7}\\
16\pi^2 \m\frac{d}{d\m} \left(
\begin{array}{c}
C_2^{\tilde H} \\
C_5^{\tilde H}
\end{array}
\right)
&=&
\left(
\begin{array}{cc}
-\frac{37}{3}g_s^2 + \frac{3}{2}y_t^2 & 2 g_s y_t\\
4 g_s y_t & -14g_s^2 + 3y_t^2
\end{array}
\right)
\left(
\begin{array}{c}
C_2^{\tilde H} \\
C_5^{\tilde H}
\end{array}
\right). \label{eq:rge_ch}
\end{eqnarray}
Eq. (\ref{eq:rge_ch}) shows, in spite of the suppression by a loop factor,
$C_5^{\tilde H}$ is significantly enhanced at low scale
because of mixing between $Q_2^{\tilde H}$ and $Q_5^{\tilde H}$.
On the other hand, $C_7^{\tilde B}$ is not enhanced
because $Q_7^{\tilde B}$ does not mix with four Fermi operators at 1 loop level. 
{
This can be seen by explicit diagram calculation.
By using Fierz identity, the four Fermi operators which contain a gluino and a Bino
can be written by,
\begin{eqnarray}
Q_{q_{L,R}}^{\tilde B} = ({\bar{\tilde B}} \g^\m \g^5 \tilde g) (\bar q \g_\m T^a P_{L,R} q).
\end{eqnarray}
At 1 loop level, only a quark loop diagram is a candidate to give mixing of
$Q_{q_{L,R}}^{\tilde B}$ and $Q_7^{\tilde B}$.
However, the quark loop contribution is proportional to
the correlation function of a quark current,
then, its transverse part is proportional to $q^2$, where $q$ is momentum of a gluon.
Therefore, when an external gluon is on-shell, its contribution is vanished.
This behavior is not guaranteed by any symmetry, therefore,
$Q_{q_{L,R}}^{\tilde B}$ and $C_7^{\tilde B}$ have mixing at 2 loop level.}

A neutralino mass eigenstate is a mixing of the Bino, Wino and Higgsino.
Then, the decay width of the gluino two body decay can be written
by the mixing matrix $N_{ij}$ and the Wilson coefficients at low scale.
{
In this paper, we neglect the radiative corrections on the neutralino and chargino mixings from the scalar particles for simplicity.
}
The decay width of the gluino two body decay is given by,
\begin{eqnarray}
\Gamma(\tilde g\to \tilde \chi_i^0 g) &=& \frac{(m_{\tilde g} - m_{\tilde \chi_i^0})^3}{2\pi m_{\tilde g}^3} \left( C_{\rm eff}^{\tilde g\to \tilde \chi_i^0 g} \right)^2, \label{eq:width}\\
C_{\rm eff}^{\tilde g\to \tilde \chi_i^0 g}(\m) &=& C_7^{\tilde B}(\m) N_{i1} + C_5^{\tilde H}(\m) N_{i4} v + \frac{g_s y_t}{8\pi^2} C_2^{\tilde H}(\m) N_{i4} v \log\frac{m_t^2}{\m^2}. \label{eq:width2}
\end{eqnarray}
Here, $v \simeq 174~{\rm GeV}$ is the SM Higgs VEV.
We have checked the above expressions are consistent with
Eq. (\ref{eq:to_bino}), (\ref{eq:to_higgsino}) at the leading order.

\subsection{Numerical evaluation of the branching fractions}
\begin{figure}[htbp]
\begin{center}
\subfigure[$m_{\tilde{q}^{1,2}} = m_{\rm stop}$]{
\includegraphics[scale=0.605]{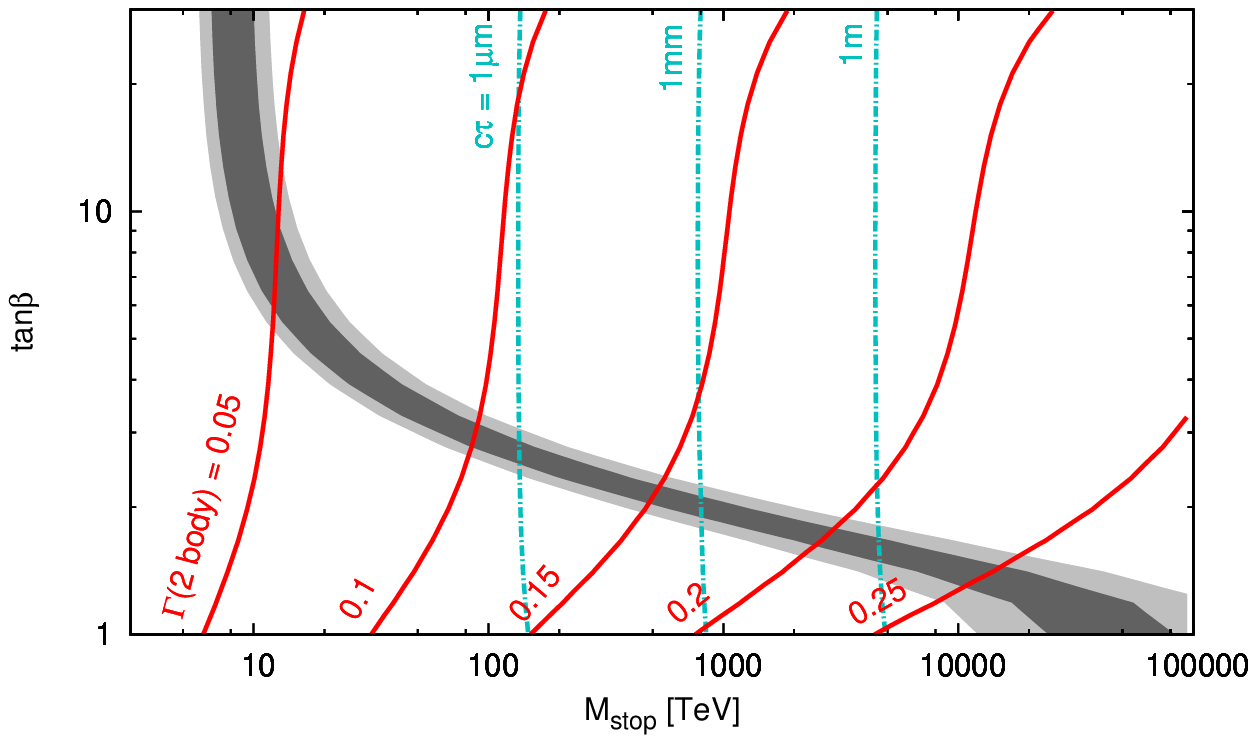}}
\subfigure[$m_{\tilde{q}^{1,2}} = 2m_{\rm stop}$]{
\includegraphics[scale=0.605]{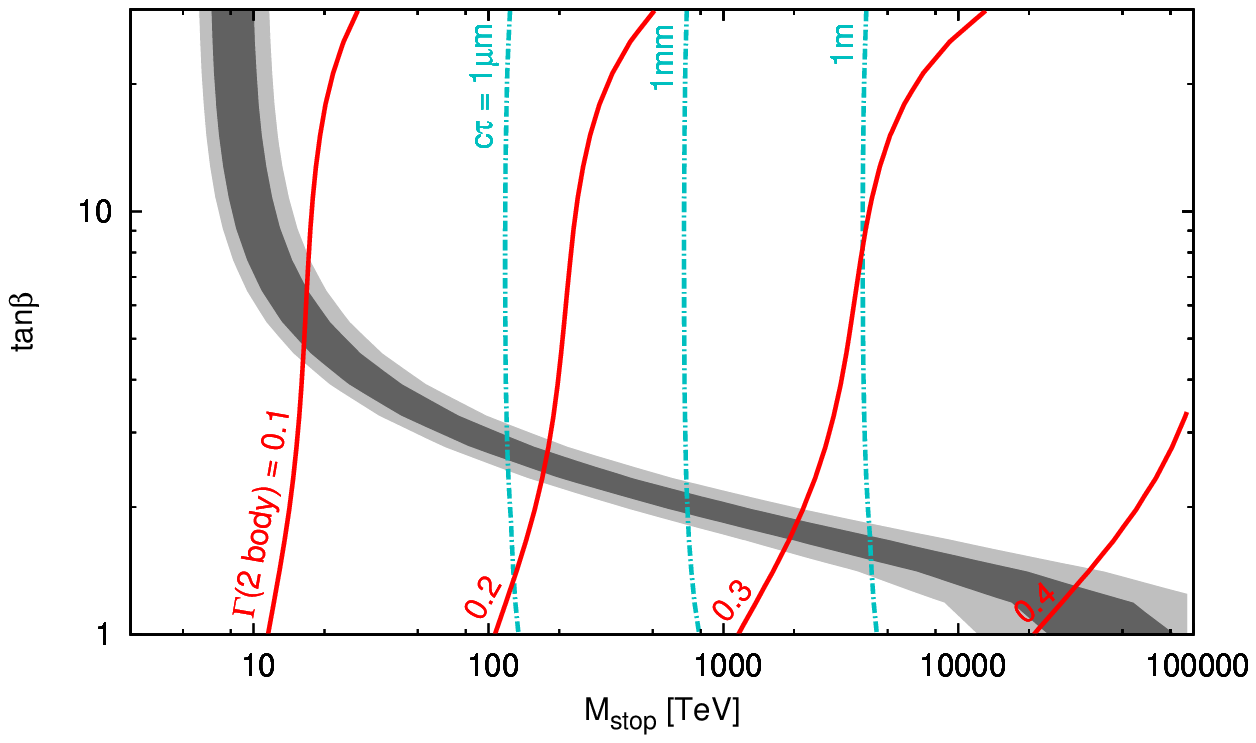}
}
\caption{
${\rm Br}({\rm 2~body})$ (solid red lines)
and $c\t_{\tilde g}$ (chain blue lines) as a function of $\tan\b$ and $m_{\rm stop}$.
In dark gray shaded region, $124~\GEV < m_h < 126~\GEV$.
The light gray region shows the uncertainty of $m_h$ due to the present error on the top quark mass.
At the stop mass scale, we take first and second generation squark mass as
$m_{\rm stop}$ (left) and $2m_{\rm stop}$ (right).
We take $M_1 = 150~\GEV$, $M_2 = 300~\GEV$, $M_3 = 1000~\GEV$, $\m = 230~\GEV$
and $173.2 \pm 0.9 ~\GEV$ \cite{Lancaster:2011wr} and $\a_s(m_Z) = 0.1184$ \cite{Bethke:2009jm}.
}\label{fig:branch}
\vspace{1cm}
\includegraphics[scale=0.8]{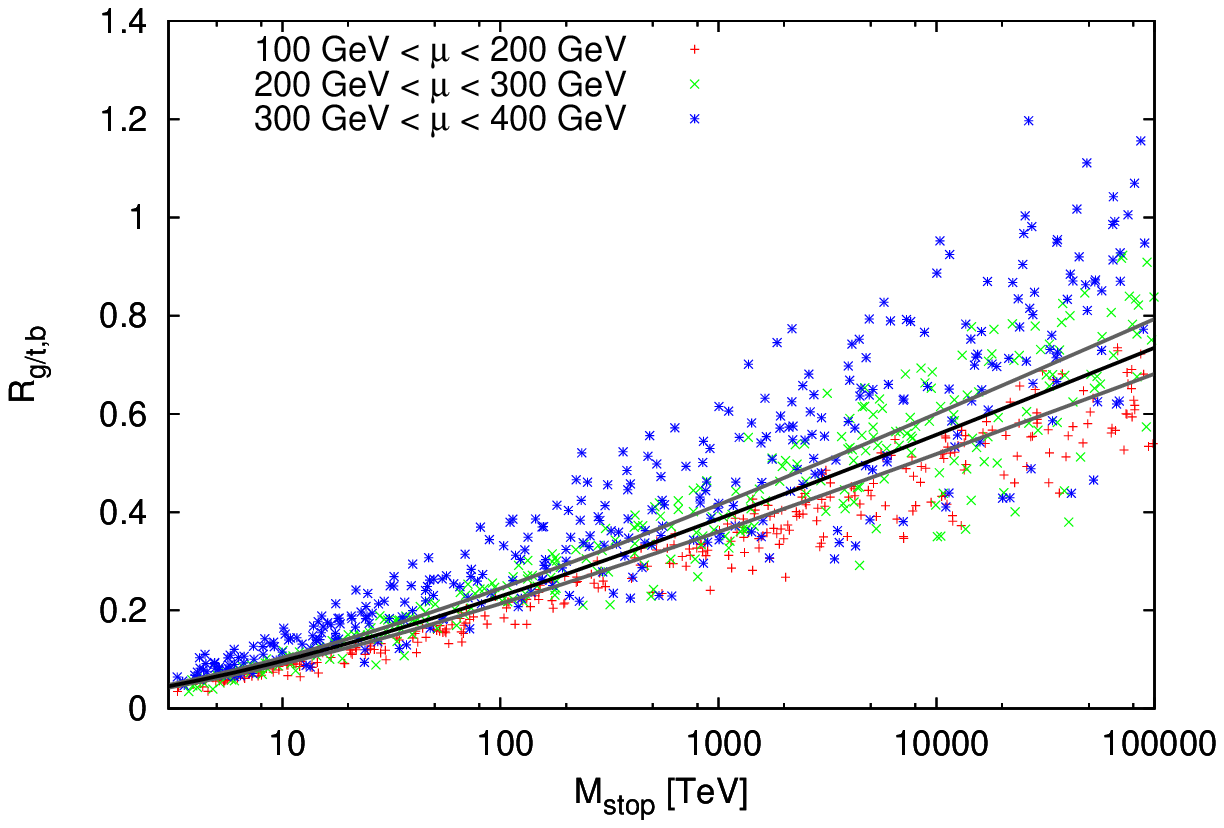}
\caption{
Third generation squark mass versus ${\cal R}_{g/t,b}\equiv\G(\tilde g\to g\tilde \chi) / \G(\tilde g\to tt/tb/bb + \tilde \chi)$.
On the black line, we take $M_1 = 150~\GEV$, $M_2 = 300~\GEV$, $\m = 230~\GEV$ and $\tan\beta = 2$,
and the gray lines show the variation if we change the parameters by 20 \%. 
{We take uniform distributions for $M_1, M_2, \tan\beta$, $m^2_{\tilde q^{1,2}}/m^2_{\tilde t_R}$, $m^2_{\tilde q^3_L}/m^2_{\tilde t_R}$ and $m^2_{\tilde b_R}/m^2_{\tilde t_R}$ at the GUT scale, while we take $\ln m_{\rm stop}$ to be uniformly distributed for $m_{\rm stop} = \sqrt{ m_{\tilde q^3_L}m_{\tilde t_R} }$.}
}\label{fig:scatter}
\end{center}
\end{figure}

\begin{figure}[htbp]
\begin{center}
\includegraphics[scale=0.8]{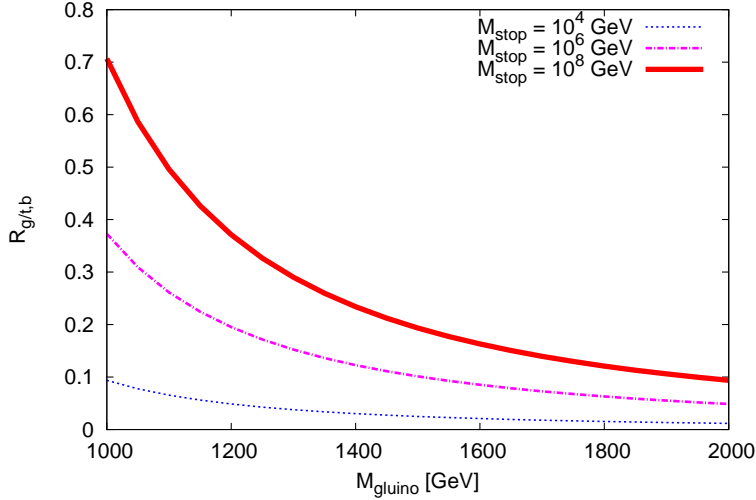}
\caption{
${\cal R}_{g/t,b}\equiv\G(\tilde g\to g\tilde \chi) / \G(\tilde g\to tt/tb/bb + \tilde \chi)$ as a function of the gluino mass.
We take the stop mass as $10^4$, $10^6$ and $10^8~\GEV$.
For other parameters, we take $M_1 = 150~\GEV$, $M_2 = 300~\GEV$, $\m = 230~\GEV$ and $\tan\beta = 2$.
We take universal squark masses at the squark mass scale.
}\label{fig:mg_dependence}
\end{center}
\end{figure}

In closing of this section, we show some numerical evaluations
of the branching fractions.
We use the RG equations given in Ref. \cite{Giudice:2011cg} for dimensionless coupling constants,
and Ref. \cite{Gambino:2005eh} for the Wilson coefficients which concern with
the two body and three body decay of the gluino.
Fig.~\ref{fig:branch} shows ${\rm Br}(\tilde g\to g \tilde\chi^0 )$, decay length $c\t_{\tilde g}$ and the lightest Higgs mass $m_h$ in the heavy scalar scenario.
We can see an enhancement of the branching fraction of the two body decay as the stop mass increases, and there exist  parameter regions consistent with  $m_h \simeq 125~\GEV$.
An important observation is that we can expect more than ${\cal O}(10)$ \% branching fraction of the two body decay when
$\tan\beta = {\cal O}(1)$.
The partial decay width of the two body decays is almost determined by the stop mass scale,
although the total decay width is sensitive to the first and second generation squark mass and so on.
Hence, to extract the stop mass scale,
${\cal R}_{g/t,b}\equiv\G(\tilde g\to g\tilde \chi) / \G(\tilde g\to tt/tb/bb + \tilde \chi)$ is
more suitable parameter than ${\rm Br}(\tilde g\to g\tilde\chi)$.
In Fig. \ref{fig:scatter}, we show a scatter plot of the stop mass
$m_{\rm stop} = \sqrt{m_{\tilde q^3_L}m_{\tilde t_R}}$ at the stop mass scale
versus ${\cal R}_{g/t,b}$.
In this figure, we take $m_{\tilde g} = 1~\TEV$ and other parameters are chosen from a parameter region:
\begin{itemize}
\item $100~\GEV < M_1,~M_2,~\m < 400~\GEV$.
\item $2 < \tan\beta < 10$.
\item $3~\TEV < m_{\rm stop} < 10^5~\TEV$.
\item $0.5 m_{\tilde t_R} < m_{\tilde q^{1,2}}, m_{\tilde q^3_L}, m_{\tilde b_R} < 2m_{\tilde t_R}$
at the Grand Unified Theory (GUT) scale.
\end{itemize}
Fig. \ref{fig:scatter} shows that
${\cal R}_{g/t,b}$ which is mainly determined by $m_{\rm stop}$,
and the information about the EW-ino masses enable us to
determine $m_{\rm stop}$ more precisely.
For larger $\mu$, the two body decay is relatively enhanced because the sizable branching fraction of three body decay into Higgsino and quarks becomes smaller.   
Fig. \ref{fig:mg_dependence} shows how ${\cal R}_{g / t,b}$ depends on the gluino mass $m_{\tilde g}$,
and this figure shows the light gluino is advantageous to enhance ${\cal R}_{g/t,b}$.
As we can see from Eq. (\ref{eq:to_higgsino}) or Eqs. (\ref{eq:width}, \ref{eq:width2}),
the enhanced part in $\G(\tilde g\to g\tilde\chi)$ needs chirality flip by the top quark mass.
Therefore, $\G(\tilde g\to g\tilde\chi)$ has a suppression factor $m_t^2 / m_{\tilde g}^2$.
Finally, we show a sample point of the  gluino decay table and mass spectrum in Table. \ref{tab:sample}.

{
In the parameter region of our interest, the branching fraction of the gluino two body decay is  ${\cal O}(10)~\%$. 
Therefore sizable amount of the gluino produced at the LHC can two-body decay.
However, it is not clear how we can extract the information of the two body decays.
This is because there are so many jets in a SUSY event and we cannot well discriminate jets from a gluino decay.
In the next section, we present a method of measuring branching fractions, and will conclude they are measurable. 
}

\begin{table}[h!]

\caption{
We show the sample point of the gluino decay table (left) and mass spectrum (right).
We take $M_1 = 150~\GEV$, $M_2 = 300~\GEV$, $\m = 230~\GEV$, $\tan\beta = 2.5$ and $m_{\tilde q} = 2 \times 10^5~\GEV$.
We assume universal squark pole masses.
${\rm Br}(\tilde g\to g\tilde \chi^0)$ = 0.123, 
${\cal R}_{g/t,b}$ = 0.459.
}\label{tab:sample}

\begin{tabular}{|c|c|c|c|c || c | c|}
\multicolumn{7}{c}{Gluino branching fractions} \\ \hline
& $g$ & $t\bar{t}$ & $b\bar{b}$ & $q\bar{q}$ & $q\bar{q}'$ & $tb$ \\ \hline
$\tilde\chi^0_1 $&  0.013 &  0.062 &  0.010  & 0.015 & - & - \\ \hline
$\tilde\chi^0_2 $&  0.029 &  0.060 &  0.0086  & 0.045 & - & - \\ \hline
$\tilde\chi^0_3 $&  0.064 &  0.00034 & 0.00030  & 0.064 & - & - \\ \hline
$\tilde\chi^0_4 $&  0.018 &  0.066 & 0.017  & 0.040 & - & - \\ \hline
$\tilde\chi^\pm_1 $&  - &  - &  -  & -  & 0.094 & 0.11 \\ \hline
$\tilde\chi^\pm_2 $&  - &  - &  -  & -  & 0.14 & 0.15 \\ \hline
\end{tabular}~~~
\begin{tabular}{|c|c|}
\hline
 & mass [GeV] \\
\hline\hline
$m_h$ & 125.1 \\
\hline\hline

$m_{\tilde \chi_1^0}$ & 128 \\
\hline
$m_{\tilde \chi_2^0}$ & 205 \\
\hline
$m_{\tilde \chi_3^0}$ & -233 \\
\hline
$m_{\tilde \chi_4^0}$ & 349 \\
\hline\hline

$m_{\tilde \chi_1^\pm}$ & 185\\
\hline
$m_{\tilde \chi_2^\pm}$ & 349\\
\hline \hline

$m_{\tilde g}$ & 1000\\
\hline

\end{tabular}

\end{table}


\section{Study of the gluino decay at the LHC}
As we discussed above, the  branching fraction of the gluino two body decay carries important information on the scalar mass scale. 
In addition, the recent Higgs observation suggests sizable branching fraction of the gluino two body decay.
In this section, we discuss the measurement of the gluino branching fractions at the LHC.
For this purpose, we adopt a simple method, using the difference of the event topology depending on the gluino branching fraction.
Then we demonstrate this method for some models.

\subsection{Basic Strategy}

First we compare the three body decay ($\tilde{g}\to q_3 q_3 \tilde{\chi}$) mediated with third-family squarks to the two body decay 
into a gluon ($\tilde{g}\to g \tilde{\chi}$).
For the former case, the gluino decays into third-family quarks, $b$ and $t$, if  small flavor violations assumed.
These $b$-jets and leptons originated from the top quark decay will leave distinct signatures from the latter two body decay case.
Therefore, it is relatively easy to discriminate the two body decay from the three body decay into the third-family quarks.

Then, let us discuss the discrimination between the two body decay and three body decay into the first and second-family quarks.
Unlike the third family case, jet flavors and leptons are not necessarily help for the discrimination.
However, the behavior of the jets $P_{\rm T}$ and missing energy will be distinguished because of the difference of kinematics of the two body and three body decay.
Naively, the three body decay case is expected to have more number of jets and each jet has less energetic compared to the two body decay case.
{
However it is difficult to directly identify the jets from the gluino decay, since there are many other sources of jets,
such as initial and final state radiations and decay from EW-inos.
Therefore we must consider a statistical way to discriminate gluino two and three body decays.}
To extract the difference, we define event cuts according to jet multiplicity and study the jet behavior together with $E_{\rm T, miss}$ in detail. 
These cuts enhance the difference of two decay patterns and also take a role of reduction of the SM background as well.

Another barometer will come from  difference of the neutralino in the gluino decay product.
As for the decay $\tilde{g} \to g \tilde\chi^0$, as discussed before, the final state  $\tilde\chi^0$ have large $\tilde H_u$ component, since
the two body decays requires the large Yukawa coupling of the top quark and the neutralino. 
On the other hand, the three body decay mediated with the first and second family squarks $\tilde{Q}_{1,2}, \tilde{d}_{R 1,2} $ and $ \tilde{u}_{R 1,2}$
responds to the gaugino components $\tilde{B}$ and $\tilde{W}$.
Therefore the species of final state neutralino from the two body and three body decay can be used as an information to distinguish the decay patterns of the gluino.

Using these differences, we can estimate the gluino branching fraction. We adopt the following method.
One of the basic strategy is owe to a simple number counting of the leptons, $b$-jets and high $P_{\rm T}$ jets.

We separate signal regions corresponding to the number of the leptons $N_\ell$, $\tau$-jets $N_\tau$ and $b$-jets $N_b$ in each event.
Here we require that the $P_{\rm T}$'s of the lepton, $\tau$ and $ b$ are larger than $30$ GeV,
50 GeV and 50 GeV, respectively.
In addition, as for the lepton numbers, we consider set of $N_{\ell}=\{0, 1, {\rm SS}2\ell, {\rm OS}2\ell, \ge 3  \}$, where 
SS2$\ell$ represents same sign 2 leptons and OS2$\ell$ opposite sign.
As for the $\tau$ and $b$-jets, $N_{\tau} =\{0,\ge 1\}$, $N_{b} =\{0,1,2,\ge 3\}$.

We also count the number of jets $N_j$ with $P_{\rm T} > 100$ GeV.
To characterize kinematical nature of the event, 
we further consider two conditions referred to as $N_j$-J1 and $N_j$-J2 for each jet multiplicity
and each event is tagged whether it satisfies these conditions or not.
After all, each event can be labeled with a set of ($N_b,N_\ell ,N_\tau, N_j$, $N_j$-J1, $N_j$-J2).\footnote{
Although the total number of signal regions is ${\cal O}(1000)$,
actually, the number of signal regions which are important for the measurement of the branching fraction is roughly 20-30.
Then, if we concentrate on specific SUSY model or mass spectrum,
we can do more sophisticated analyses.
In this paper, we propose a method which can be used for generic heavy scalar models.
}
We always require  $E_{\rm T, miss}>300$ GeV as a basic cut.

For multi-$b$-jets and leptons modes, this basic cut and requirement of multi 100 GeV $P_{\rm T}$ jets reduce the
SM background effectively.
However, in the case of less $b$-jets and leptons events, the basic cut is not adequate for the SM background reduction and also is difficult to discriminate two body decay and three body decay.
Then the conditions $N_j$-J1 and  $N_j$-J2 play an important role. 
They are defined as follows. 

\begin{description}
\item[2 jets mode $N_j=2$]\mbox{}\\
$\Delta \phi(j_{1,2}, \vec{E}_{\rm T, miss})>0.4$ rad and 
 \begin{description}
 \item[2J1:] $P_{j1,{\rm T}}>500$ GeV,  $P_{j2,{\rm T}}>250$ GeV, $E_{\rm T,miss}>600$ GeV, $M_{\rm T2}>650$ GeV
 \item[2J2:] $P_{j2,{\rm T}}>150$ GeV, $M_{\rm eff,2}>1200$ GeV,  $E_{\rm T,miss}> \max(0.25M_{\rm eff,2}, 400 ~{\rm GeV})$.
 \end{description}
\item[3 jets mode $N_j=3$]\mbox{}\\
$\Delta \phi(j_{1,2,3}, \vec{E}_{\rm T, miss})>0.4$ rad and 
 \begin{description}
 \item[3J1:] $P_{j1,{\rm T}}>350$ GeV,  $P_{j2,{\rm T}}>300$ GeV, $M_{\rm eff,3}>1200$ GeV
 \item[3J2:] $P_{j2{\rm T}}>150$ GeV, $M_{\rm eff,3}>1500$ GeV,   $E_{\rm T,miss}> 0.35M_{\rm eff,3}$.
 \end{description}
 
\item[4 jets mode $N_j=4$]\mbox{}\\
$\Delta \phi(j_{1,2,3,4}, \vec{E}_{\rm T, miss})>0.4$ rad and 
 \begin{description}
 \item[4J1:] $P_{j2,{\rm T}}>250$ GeV,  $P_{j4,{\rm T}}>200$ GeV, $M_{\rm eff,4}>1400$ GeV and  $E_{\rm T,miss}> 0.1M_{\rm eff,4}$.
 \item[4J2:] $P_{j2{\rm T}}>150$ GeV, $M_{\rm eff,4}>1400$ GeV,   $E_{\rm T,miss}> \max(0.35M_{\rm eff,4}, 600 ~{\rm GeV})$.
 \end{description}
 
\item[5 jets mode $N_j=5$]\mbox{}\\
$\Delta \phi(j_{1,2,3,4}, \vec{E}_{\rm T, miss})>0.4$ rad and 
 \begin{description}
 \item[5J1:] $P_{j1,{\rm T}}>400$ GeV,  $P_{j2,{\rm T}}>150$ GeV, $M_{\rm eff,5}>1400$ GeV, \\
  $E_{\rm T,miss}> \max(0.25M_{\rm eff,5}, 700 ~{\rm GeV})$.
 \item[5J2:] $P_{j2{\rm T}}>300$ GeV,  $P_{j3,{\rm T}}>150$ GeV,  $E_{\rm T,miss}>0.1 M_{\rm eff,5}$.
 \end{description} 
 
\item[6 and more jets  mode $N_j \ge 6$]\mbox{}\\
$\Delta \phi(j_{1,2,3,4}, \vec{E}_{\rm T, miss})>0.4$ rad and 
 \begin{description}
 \item[6J1:] $P_{j1,{\rm T}}>200$ GeV,  $P_{j6,{\rm T}}>150$ GeV, $M_{\rm eff, all}>1400$ GeV.
 \item[6J2:] $P_{j3{\rm T}}>150$ GeV, $M_{\rm eff,all}>1400$ GeV.
 \end{description}  
 
\end{description}
Here $P_{jn, {\rm T}}$ is the $n$-th highest $P_{\rm T}$ of jets and $M_{{\rm eff}, i}\equiv \sum_{1\le n\le i} P_{jn,{\rm T}}+E_{\rm T,miss} +\sum_{\rm leptons} P_{i,{\rm T}}$.
{
In signal regions corresponding to number of jets, 
we employ $\Delta \phi$ cut which significantly reduces the QCD background. }
Note that the cuts imposed above are pretty normal, and actually similar cuts are already used in the current analysis \cite{ATLAS-CONF-2012-033}.   
By using the number of events after each cut, we can confirm the above qualitative arguments in a quantitative form.

Then let us consider the determination of the branching fraction.
We use the maximum likelihood estimation (MLE).
We consider the following likelihood $\cal L$,
\beq
{\cal L} = \prod_{{\rm modes}}
{\rm Prob}(N_{\rm mode} |\{{ B}_{i},\sigma_{ \tilde{g}\tilde{g}}\} ),
\eeq
where ${B}_i$ is a set of branching fractions of the gluino, $\sigma_{ \tilde{g}\tilde{g}}$ is the gluino production cross section and
$N_{\rm mode}$ is the number of the events characterized with ($N_b,N_\ell ,N_\tau, N_j$, $N_j$-J1, $N_j$-J2).
We assume ${\rm Prob}$ is a convolution of Poisson distribution from the event number and Gaussian distribution from systematic uncertainties of
the acceptance of the SUSY signals and the number of the SM backgrounds.
In this study, we fix the parameter such as the gluino mass except for the gluino branching fractions and the cross section.
As for the systematic uncertainties of the number of the SM $\delta_B$, we assume $\delta_B =20 $ \% 
{\cite{Aad:2009wy}. 
This error estimation may be too optimistic for the QCD background.
However in the signal regions which is significant for measurement of the gluino branching fractions,
the rate of the QCD background is sub-dominant compared to  $Wj,Zj$ and $t\bar{t}$ backgrounds 
and the final result is insensitive to the detailed value of uncertainty of the QCD backgrounds.
} 
For acceptances of the SUSY events  ($\delta_S$), we set  $\delta_S = 20 $ \%.
In the above set-up, we consider the minimization of $-2 \ln({\cal L})$ by varying
$\{{ B}_{i},\sigma_{ \tilde{g}\tilde{g}}\}$
and estimate best fitting parameters and
their errors as the usual MLE method.

In the following, we show the result of Monte-Carlo simulation.
To estimate the SM background, 
we have
used the programs MC@NLO \cite{Frixione:2002ik}
(for $t\bar{t},WW,WZ$ and $ZZ$),
Alpgen \cite{Mangano:2002ea} (for $Wj,Zj$ and $W/Z+b\bar{b}/t\bar{t}$) and
Pythia 6 \cite{Sjostrand:2006za} (for QCD jets).
As for the detector simulation, we used the AcerDet \cite{RichterWas:2002ch}.
For the isolation criteria for leptons and photons,
we used the default setting of the AcerDet.
In our study, we are interested in the signals with leptons, $\tau$ jets and $b$-jets.
The detection efficiency and misidentification of these particles are important in estimating both the
signal events and background events.
In the following simulation, we include the misidentification and fake rates of the leptons, $\tau$, $b$-jets and light jets, following 
Ref. \cite{Aad:2009wy}.
As for the MSSM mass spectrum, we have used the program ISAJET 7.82  \cite{ISAJET} modified for the scalar decoupling scenario.
The SUSY events are also generated with the program Pythia \cite{Sjostrand:2006za}.
The NLO production cross section of the gluino production is estimated with the program Prospino2 \cite{Beenakker:1996ed}.

\subsection{Examples}
We apply the above method on three models as examples, and see how accurate we can determine the branching fraction of the gluino two body decay which is sensitive to the mass scale of squarks. 
We consider a simplified model which contains only a gluino, a neutralino and a chargino,
the split SUSY model and the anomaly mediation model.

\subsubsection{Simplified Model}
\subsubsection*{Setup}
Let us consider only three particles are relevant: gluino $\tilde{g}$ (1000 GeV), a light neutralino (100 GeV) $\tilde\chi^0$ and 
a chargino $\tilde\chi^\pm$ slightly heavier than the neutralino (by 100 MeV).
This simplified set-up is very useful to see the how well the above method works.
We set the chargino decay $\tilde\chi^\pm \to e\nu \tilde\chi^0$, whose
decay products are irrelevant for the collider study. 
The collider signature can be characterized by the branching fractions of the gluino:
$\tilde{g} \to g\tilde\chi^0, t\bar{t} \tilde\chi^0, t{b} \tilde\chi^\pm,  b\bar{b} \tilde\chi^0$ and $  q\bar{q} \tilde\chi^0$.
Thus, we vary branching fractions of these decay modes, $B_i$, as well as the production cross section of the gluino-pair, $\sigma_{\tilde{g}\tilde{g}}$. 

\subsubsection*{Measurement of the Branching Fractions}
We apply the strategy discussed in Section 3.1 to this simplified model, 
and determine a set of \{$B_i, \sigma_{\tilde{g}\tilde{g}}$\} based on the MLE. 
{
To see the difference of the event topology intuitively, it is useful to use the significance variable $Z$, 
which represents the deviation from the SM background \cite{Asai:2012hb}.
Roughly speaking, this variable $Z$ is given by ratio of the number of SUSY events to that of the fluctuations of the SM background,
$Z \sim N_s/\Delta N_b$, where $N_s$ is the number of SUSY events and $\Delta N_b$ is statistical and systematic uncertainties for 
the number of the SM background.
The definition of $Z$ is as follows: 
Given the expected number of the signal events $N_s$ and the background events $N_b$ with the uncertainty $\delta N_b$, 
the significance is given by calculating the convolution of the Poisson distribution with some ``posterior'' distribution function.
As the posterior distribution, we take the gamma distribution as suggested in Ref.~\cite{Linnemann}. 
The resulting significance $Z$ is given by \cite{Linnemann} with
	\begin{eqnarray}
	Z=\sqrt{2} {\rm erf}^{-1}(1-2p_B),
	\end{eqnarray}
with
	\begin{eqnarray}
	p_B=\frac{B \left(N_s+N_b, 1+N_b^2/\delta N_b^2, \delta N_b^2/(N_b+\delta N_b^2)\right)}{B(N_s+N_b, 1+N_b^2/\delta N_b^2)}, \label{pB}
	\end{eqnarray}
where erf$^{-1}$ is the inverse error function and
	\begin{eqnarray}
	B(a,b,x)= \int_0^x dt t^{a-1}(1-t)^{b-1}
	\end{eqnarray}
is the incomplete beta function. 
If we take the limit $\delta N_b \to 0$, the Eq.~(\ref{pB}) reduces to the probability in the usual Poisson distribution. 
In the case of smaller background $N_b < 0.1$, we conservatively take the $N_s$ as the significance.
}

\begin{figure}\begin{center}
\includegraphics[clip, height=0.5\columnwidth]{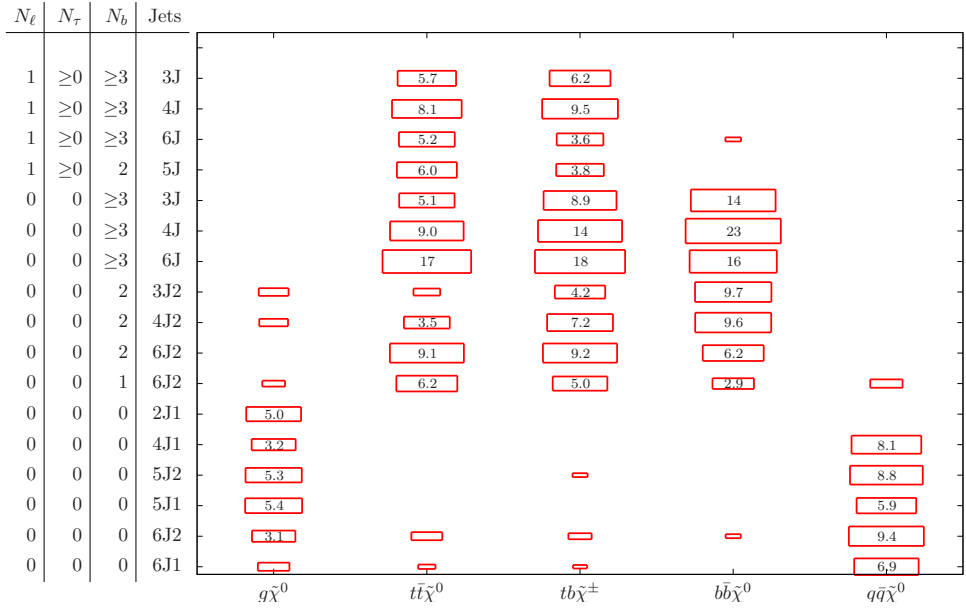}
\caption{Significance variable $Z$ for each mode in the simplified model.
{
$Z$ represents the signal strength compared to the SM backgrounds.
See text for detail.
}
In the row of Jets, $N$J  represents the number of the  jets  with $P_{\rm T}>100$ GeV and
$N$J1,2 represent the additional conditions in Sec 3.1.
}
\label{fig:sig_simp}
\end{center}
\end{figure}

In Fig. \ref{fig:sig_simp}, we show the significance variables $Z$'s of some selected cuts which is important for the parameter estimation.
Here we assume an integrated luminosity of 5 fb$^{-1}$ at
$\sqrt{s} = 14$ TeV LHC run.
As expected, the gluino decay into the third-family quarks provides very distinct signatures from the gluino two body decay.
As for the difference of  decay into $q\bar{q}$ and $g$, we can see the clear difference when we compare signal regions of higher and lower jet multiplicities.  
The the decay mode of $\tilde{g}\to q\bar{q}\tilde{\chi}^0$ shows large $Z$ for large jet multiplicity modes.

\begin{figure}[t!]
 \begin{center}
 
\includegraphics[clip, width=0.5\columnwidth]{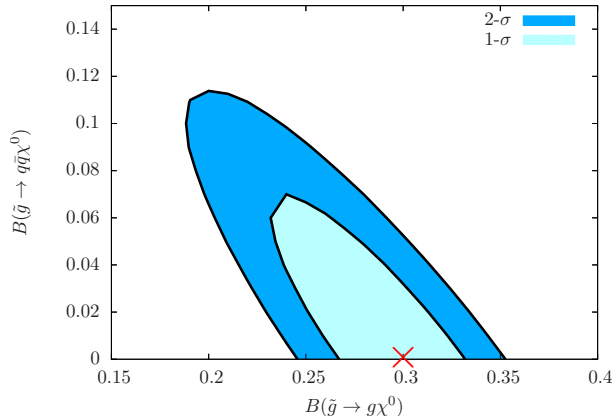}
\caption{1-$\sigma$ and 2-$\sigma$ contour lines of the estimations 
of $B(\tilde{g}\ \to g\tilde\chi)$ and $B(\tilde{g}\ \to t\bar{t}\tilde\chi)$
at the sample point in the simplified model.
The point $\times$ represents the fiducial value.
}
\label{fig:para_S}
\end{center}
\end{figure}

In Fig. \ref{fig:para_S}, we show 1 and 2-$\sigma$ contour lines of expected parameter estimation for combination of
$B(\tilde{g}\ \to g\tilde\chi)$ and $B(\tilde{g}\ \to qq\tilde\chi)$  with
an integrated luminosity 10 fb${}^{-1}$ at $\sqrt{s}=14$ TeV.
Here we assume  that $B(\tilde{g}\ \to g\tilde\chi)=0.3$, $B(\tilde{g}\ \to t\bar{t}\tilde\chi)=0.7$ and others 0 as fiducial values.

There is a direction, $B(\tilde{g}\ \to g\tilde\chi) + B(\tilde{g}\ \to qq\tilde\chi) \simeq 0.3$,  
in which it is slightly difficult to determine each branching fraction 
since both of them do not generate $b$-jets or leptons. 

In Table \ref{tb:simp}, we show the expected accuracy of determination of the branching fraction of two body decay for some sets of the parameters.
In the simplified model, we can see the branching fraction of two body decay  can be determined to accuracy of 0.02 - 0.05.

\begin{table}[htbp]
\caption{Expected accuracy of two body branching fraction for the simplified model.}
\label{tb:simp}

\begin{center}
  \begin{tabular}{ |c|c|c| }
    \hline
    Fiducial Values &  $\cal L$ [${\rm fb}^{-1}$]  & $1 \sigma$ estimation \\ \hline
    $(B(g), B(t\bar{t})) =(0.3, 0.7)$ & 10 & (0.26, 0.32) \\ \hline
     $(B(g), B(t\bar{t})) =(0.3, 0.7)$ & 25 & (0.27, 0.31) \\ \hline
        $(B(g), B(q\bar{q})) =(0.2, 0.8)$ & 10 & (0.15, 0.24) \\ \hline  
                $(B(g), B(q\bar{q})) =(0.2, 0.8)$ & 25 & (0.17, 0.23) \\ \hline  

      \end{tabular}
      \end{center}
\end{table}

\subsubsection{Split SUSY}
\subsubsection*{Setup}
Then let us consider a more realistic example.
Here, we discuss the scalar decoupling scenario.
At the low energy, there are the gauginos ($\tilde{B}$, $\tilde{W}$ and $\tilde{g}$) and 
the Higgsinos ($\tilde{H}_u, \tilde{H}_d$), which are
relevant for the LHC study.
We assume the Higgsino mass is the same order as the Bino and Wino masses,
and then, EW-inos and Higgsinos are mixed with each other and form the mass-eigenstates: neutralinos and charginos.\footnote{
To be precise, the mixing parameters in the split model are different from 
ones of the usual MSSM scenario, because of the radiative corrections due to
the large mass hierarchy between the squarks and the gauginos.
However, we neglect this difference for simplicity.
}
We assume that the gauginos masses and the mixing parameters are well-known.
This assumption would be justified by considering that the neutralino can be the dark matter.
For example, the abundance, the cross section to nucleon and mass and/or cosmic ray signatures at future experiments
provide information on the EW-ino sector.
Kinematic information at the LHC  also gives a large number of clues to determine this SUSY EW-ino sector \cite{Kersting:2008qn, Turlay:2010bk}.
Or, in an extreme argument, an $e^+e^-$ linear collider such as the ILC can provide 
the precise information on the neutralino and chargino sector \cite{Kilian:2004uj}.
As for the neutralinos and charginos decay, the contribution from the heavy scalars are negligible when the scalar mass are much heavier than TeV scale.
\footnote{In this set-up, the dark matter abundance $\Omega h^2$ is around $0.1$.
If the scalar scale is around  1000 TeV, the SM-like Higgs mass is around 125 GeV.
Although the latest result of XENON100 collaboration excludes this point \cite{:2012nq},
the present analysis does not depend on details of the nature of the dark matter.
}
In this case, the branching fraction of the gluino two body decay is a few ten percent, if universal scalar masses assumed.

Then we consider the gluino decay.
As discussed in the previous section,
the gluino decays are induced by the dimension 5 and 6 operators.
However, there are a lot of the operators which induce the gluino decay,
and then, it makes the analysis very complicated.
To make the analysis easier, we take two simplifications.
First simplification is a limitation of the operators.
We only introduce the following seven operators:
\begin{eqnarray}
{\cal L}_{\rm gluino} &=&
 C_{q_R q_R} (\bar{\tilde{B}} \g^\m \g^5 \tilde{g} )( \bar q \g^\m P_R q) 
~+~ C_{t_R t_R} (\bar{\tilde{B}} \g^\m \g^5 \tilde{g} )( \bar t \g^\m P_R t )
 ~+~ C_{b_R b_R} (\bar{\tilde{B}} \g^\m \g^5 \tilde{g} )( \bar b \g^\m P_R b  ) \nonumber\\
&&
~~+~~ C_{q_L q_L} (\bar{\tilde{W}} \g^\m \g^5 \tilde{g} )( \bar q \g^\m P_L q )
~~+~~ C_{Q^3 Q^3} (\bar{\tilde{W}} \g^\m \g^5 \tilde{g} )( \bar Q_3 \g^\m P_L Q_3 ) \nonumber\\
&&
~~+~~ C_{Q^3 t_R}  (\bar{\tilde{H}}_u \tilde{g} )( \bar t P_L Q_3 )
~~+~~ C_{g}~ G_{\m\n} \bar{\tilde{H}}_u^0 \sigma^{\m\n} \tilde{g}, \label{eq:gluino_eff}
\end{eqnarray}
where $q$ collectively represents first and second generation quarks.
Given the seven $C$'s in Eq. (\ref{eq:gluino_eff}), $M_1$, $M_2$, $\m$, $\tan\b$ and $m_{\tilde g}$, we can calculate the decay width of each decay mode.
For example, let us consider $\G(\tilde g \to t\bar t \tilde\chi^0)$.
This decay width is determined by
$C_{t_R t_R}$, $C_{Q^3 Q^3}$ and $C_{Q^3 t_R}$, such as,
\begin{eqnarray}
\G(\tilde g \to t\bar t \tilde\chi^0) &=&
C_{t_R t_R}^2 {\cal A}_{t_R t_R}
~+~ C_{Q^3 Q^3}^2 {\cal A}_{Q^3 Q^3}
~+~ C_{Q^3 t_R}^2 {\cal A}_{Q^3 t_R} \label{eq:decaywidth}\\
&& + C_{t_Rt_R} C_{Q^3 Q^3} {\cal B}_{t_Rt_R,Q^3 Q^3}
~+~ C_{t_Rt_R} C_{Q^3 t_R} {\cal B}_{t_Rt_R,Q^3t_R}
~+~ C_{Q^3 Q^3} C_{Q^3 t_R} {\cal B}_{Q^3 Q^3,Q^3 t_R}, \nonumber
\end{eqnarray}
where ${\cal A}$'s and ${\cal B}$'s are functions of the masses of
the quarks, neutralinos and charginos.
The first line of the above equation shows a linear combination of the contribution
in case that only one operator is switched on and the others off.
The second line shows interference effects between the operators
due to the top quark mass and the EW-ino mixings.
The first line can be calculated easily
by a linear combination of the contribution from each operators,
although the second line can not.
Hence, we take a second simplification.
We drop the interference terms between different operators,
i.e., the second line in Eq (\ref{eq:decaywidth}).
Such a simplification is justified if the gluino mass is much larger
than the top quark mass and the EW-ino mixing terms,
and it makes the gluino width into 
a simple linear combination of the decay width in case that
only one operator is introduced.
Similarly, the branching fraction can be written by a linear combination.
For example,
if we denote the branching fractions ${\rm Br}_i(\tilde g \to t\bar t\tilde \chi^0)$ in case that only $C_i$ is switched on,
the general branching fractions is given by,
\begin{eqnarray}
{\rm Br}(\tilde g \to t\bar t\tilde \chi^0) =
\sum_i B_i {\rm Br}_i(\tilde g \to t\bar t\tilde \chi^0),
\end{eqnarray}
where $\{B_i\}=\{\
B^{2}_{g}, B^{3}_{t_RQ_3}, B^3_{t_R t_R}, B^3_{Q_3 Q_3}, B^3_{b_R b_R}, B^3_{q_L q_L}, B^3_{q_R q_R}
\}$ with $\sum_i B_i = 1$.
$B_i$'s have common value for all decay modes.
$B_i$'s are weights in a linear combination,
and they are determined as the functions of the Wilson coefficients $C_i$.
$B^{2}_{g}$ corresponds to the branching fraction of the two body decay of the gluino,
because the only operator $C_{g}~ G_{\m\n} \bar{\tilde{H}}_u^0 \sigma^{\m\n} \tilde{g}$ in Eq. (\ref{eq:gluino_eff}) is able to induce the two body decay.
From now on, we tackle with an optimization problem of $B_i$ instead of $C_i$.
Here, we set $m_{\tilde{g}} = 1000$ GeV,  $M_2 = 300$ GeV,  $M_1 = 150$ GeV,
$\mu = 230$ GeV and $\tan\beta=2$ as a sample point.

\subsubsection*{Measurement of the Branching Fractions}
\begin{figure}\begin{center}
\includegraphics[clip, height=0.45\columnwidth]{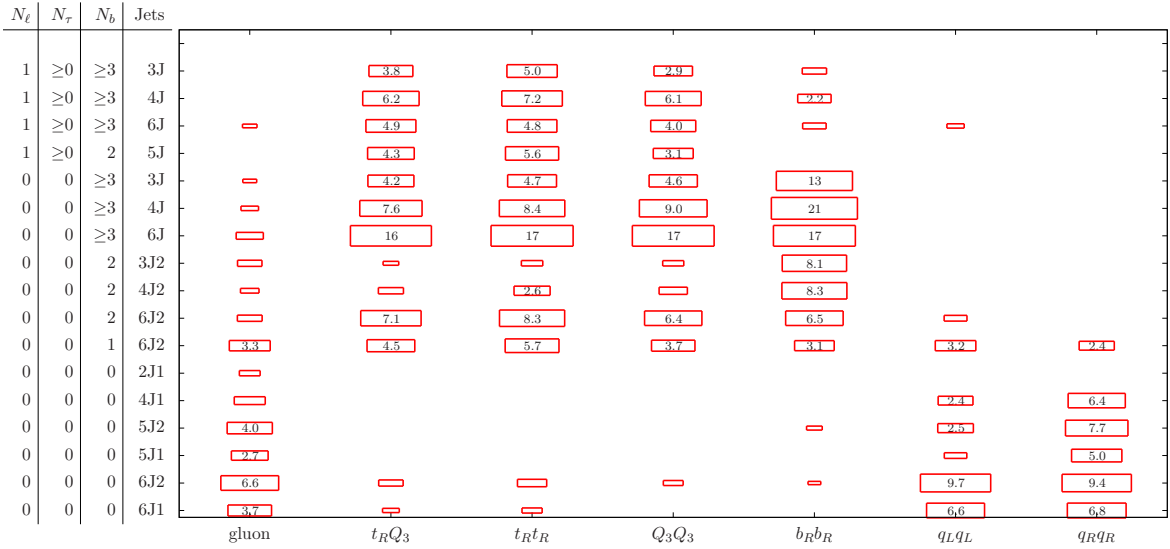}
\caption{Significance variable $Z$'s for each mode in the split SUSY model.
Each mode $i$ corresponds to the case that the coefficient $C_i$ in Eq. (\ref{eq:gluino_eff}) is dominated.
}
\label{fig:sig_split}
\end{center}
\end{figure}

Then let us move to the parameter estimation.
Under the above assumptions, we estimated accuracy in determining the gluino branching fractions at the LHC, using the MLE.
In Fig. \ref{fig:sig_split}, we show significance variables $Z$'s for some cuts for ${\cal L}=5 ~{\rm fb}^{-1}$ at $\sqrt{s}=14$ TeV.
As in the case of the simplified model, the two body decay mode of the gluino has a smaller $Z$ for high jet multiplicity modes. As expected,  the three body decays into third family quarks are distinguished by the number of $b$-jets and/or leptons.  

\begin{figure}
 \begin{center}
\includegraphics[clip, width=0.5\columnwidth]{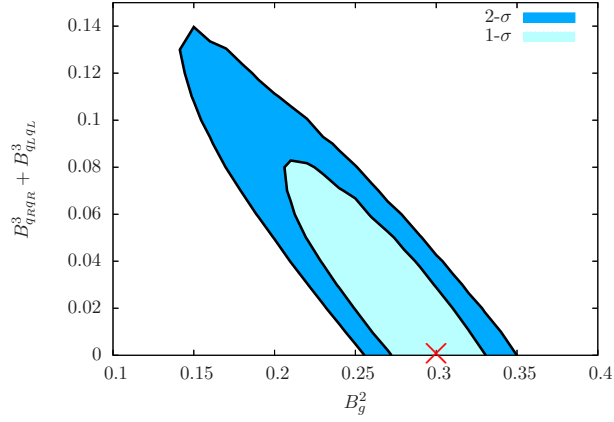}
\caption{1-$\sigma$ and 2-$\sigma$ contour lines of the estimations at the sample point in the split SUSY model in which
$B(\tilde{g}\ \to g\tilde\chi)=0.3$ and $B(\tilde{g}\ \to t\bar{t}\tilde\chi)=0.7$.
The point $\times$ represents the fiducial value.
}
\label{fig:para_R}
\end{center}
\end{figure}

In Fig. \ref{fig:para_R}, we show 1 and 2-$\sigma$ contour lines of  expected parameter estimation for 
$B^{2}_{g}$ and $B^3_{q_R q_R}+ B^3_{q_L q_L}$.
Here we assume  that $B(\tilde{g}\ \to g\tilde\chi)=0.3$, $B(\tilde{g}\ \to t\bar{t}\tilde\chi)=0.7$ and others 0 and that
an integrated luminosity 25 fb${}^{-1}$ at $\sqrt{s}=14$ TeV.
As in the case of the simplified model,
we can see there is a direction, $B^{2}_{g} + B^3_{q_R q_R}+ B^3_{q_L q_L} \simeq 0.3$,
in which it is slightly difficult to determine each branching fraction.

In Tab. \ref{tb:split}, we show the expected accuracy of determination of the branching fraction of two body decay.
We also show the case that some of the explanatory variables are constrained.
As expected, the gluino three body decay into the first and second-family quarks mimics the signals
of the gluino two body decay.
With ad hoc assumption of the absence of such three body decay, a large improvement of the parameter estimation is possible.
%
Once we get the information about the gluino decay branching fraction,
we can get the stop mass from ${\cal R}_{g/t,b}$ defined in the previous section.
For example, if we take $M_1 = 150~\GEV$, $M_2 = 300~\GEV$, $\m = 230~\GEV$ and $\tan\b = 2$,
we can get the stop mass from the black line in Fig. \ref{fig:scatter}.
From Tab. \ref{tb:split}, 
we can get an uncertainty $\d {\cal R}_{g/t,b}$ is roughly given by 0.05 at $50~{\rm fb}^{-1}$.
This uncertainty corresponds to the stop mass uncertainty $\d \log m_{\rm stop} \simeq 0.7$,
then, roughly, we can get the stop mass with the uncertainty of the factor 2.

\begin{table}

\caption{Expected accuracy of determination of the two body branching fraction of the gluino in  the split model.}
\label{tb:split}
\begin{center}
  \begin{tabular}{ |c|c|c|c| }
    \hline
    Fiducial & $\cal L$  [${\rm fb}^{-1}]$& $1$-$\sigma$ estimation  & Condition \\ \hline
    $(B_g^2, B^3_{t_RQ_3})=(0.3, 0.7)$ & 25 & (0.23, 0.32)& \\ \hline
    $(B_g^2, B^3_{t_RQ_3})=(0.2, 0.8)$& 25 & (0.14, 0.22)& \\ \hline    
    $(B_g^2, B^3_{t_RQ_3})=(0.3, 0.7)$& 50 & (0.25, 0.31)& \\ \hline
    $(B_g^2, B^3_{t_RQ_3})=(0.2, 0.8)$& 50 & (0.16, 0.21)& \\ \hline    
    $(B_g^2, B^3_{t_RQ_3})=(0.3, 0.7)$& 25 & (0.28, 0.32)& $ B^3_{q_R q_R}= B^3_{q_L q_L}=0$ \\ \hline   
     $(B_g^2, B^3_{t_RQ_3}, B^3_{q_Lq_L})=(0.2, 0.75, 0.05)$& 25 & (0.14, 0.24)& \\ \hline    
     $(B_g^2, B^3_{t_RQ_3}, B^3_{q_Lq_L})=(0.2, 0.75, 0.05)$& 50 & (0.16, 0.23)& \\ \hline   
      \end{tabular}
      \end{center}
\end{table}

Let us comment on the uncertainties from the EW-ino sector.
In this study, we assume that the parameters of this sector are precisely determined.
However, 
this is not a crucial assumption at the determination of the branching fractions.
The present method mainly relies on the the third family quarks and high $P_{\rm T}$ jets from the gluino decay and
not on the detail of the neutralino and chargino sector so much.
Actually, we have checked that ${\cal O}(10) $\% uncertainties of the parameter of the EW-ino sectors
do not change the result significantly.
On the other hand,
at the determination of the stop mass from ${\cal R}_{g/t,b}$,
Fig. \ref{fig:scatter} shows us
that ${\cal O}(10) $\% uncertainty about the parameters in the EW-ino sector
introduces the factor 2-3 uncertainty about the stop mass.
Then, we do not need the detail of the EW-ino parameters
to get the order of the stop mass,
however, the precise determination of the stop mass requires
both the EW-ino sector information and more large luminosity.
The present method is robust but crude, and there can be room for improvement.
When addressing this improvement, we must deal with uncertainties from EW-ino sectors, which is
out of the scope of this paper.

\subsubsection{Anomaly Mediation}
\subsubsection*{Setup}
Another important scenario of the heavy scalar scenario is anomaly mediation models.
In this model, the gaugino masses are generated from heavy gravitino mass thorough a conformal anomaly.
The gaugino mass from the gravity mediation can be suppressed, if appropriate symmetry assumed.
In this case, the Wino is the lightest supersymmetric particle (LSP).
On the other hand, it is difficult to forbid the scalar mass generation from the gravity mediation effects by
using some symmetry.
Therefore a natural prediction of the AMSB model is heavier 
scalar mass around the gravitino mass (${\cal O}(100)\times m_{\rm gaugino}$) or more.
Unlike the split SUSY case, we assume that the Higgsino mass is as heavy as the scalar particles.

In the case of the AMSB model, the charged Wino can be long-lived because of small mass difference between charged and neutralino Winos.
This charged track can be detected at the LHC if a large number of SUSY particles have adequately large cross section, which
is provides an important test of the AMSB model  \cite{Ibe:2006de,Asai:2008sk,Asai:2008im}.
However, the decay length of the charged Wino is typically small ${\cal O}(1-10)$ cm and
dominant signals of the AMSB models are just typical SUSY signals, i.e, large $P_{\rm T}$ jets and large missing energy.
Here we focus on this dominant AMSB signals.

Let us consider the gluino decay in the AMSB model.
The gluino two body decay into a neutralino and a gluon mainly comes from the Higgsino component.
Thus in this AMSB case, decay $\tilde{g} \to g \tilde{\chi}_0$ is significantly suppressed.
Therefore the AMSB predicts absence of the gluino two body decay and its confirmation provides an important test for the AMSB model.
Here we discuss how well the gluino two body decay can be constrained.
The gluino decay can be essentially parametrised by following branching fractions:
$ \tilde{g} \to g\tilde{W}^0, g\tilde{B}^0,
t\bar{t}\tilde{W}^0,  b\bar{b}\tilde{W}^0,  tb\tilde{W}^\pm,
 qq\tilde{W}, t\bar{t}\tilde{B}^0,  b\bar{b}\tilde{B}^0,q\bar{q}\tilde{B}^0$.
As a sample point, we assume $m_{\tilde{g}}=1000$ GeV, $m_{\tilde{B}} = 450$ GeV and  $m_{\tilde{W}} = 150$ GeV.

\subsubsection*{Measurement of the Branching Fractions}
\begin{figure}\begin{center}
\includegraphics[clip, height=0.42\columnwidth]{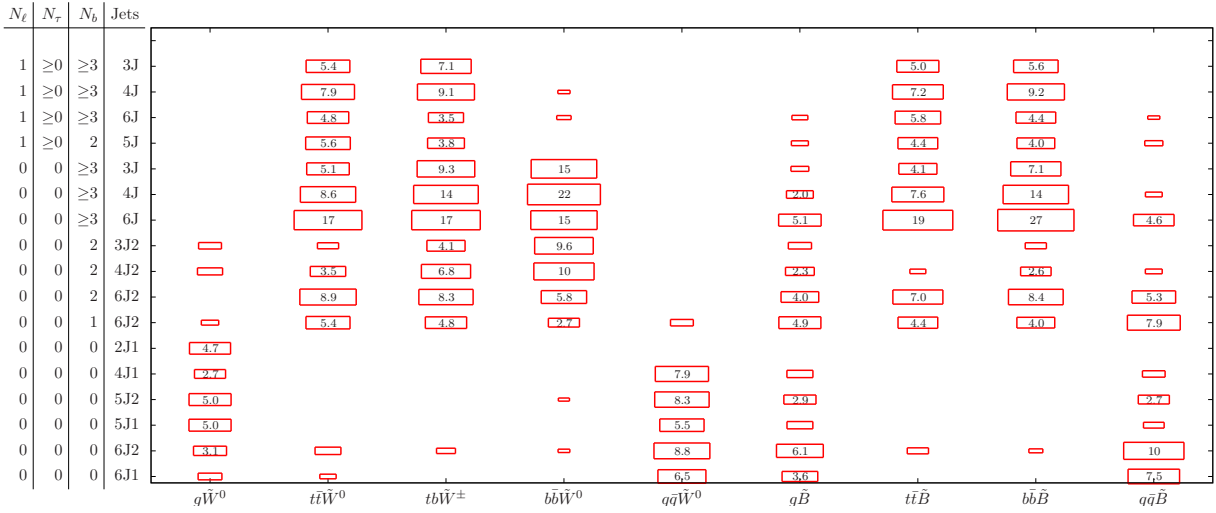}
\caption{Significance variable $Z$'s for each mode in the AMSB model.
}
\label{fig:sig_amsb}
\end{center}
\end{figure}

Using the method described previously, we constrain the gluino two body branching fraction.
In Fig. \ref{fig:sig_amsb}, we show significance variables $Z$'s for some cuts for ${\cal L}=5 ~{\rm fb}^{-1}$ at $\sqrt{s}=14$ TeV.
As in the case of the simplified model, gluino two body mode have smaller $Z$ for large jets multiplicities modes.
In Table \ref{tb:amsb}, we show the expected upper bound on the two body branching fraction.
In this case, we can see that the gluino decay width well constrained, which gives good test for the AMSB scenario.

\begin{table}

\caption{Expected 1-$\sigma$ upper bound on the
gluino two body branching fractions
at the example point of the AMSB model.}
\label{tb:amsb}
\begin{center}
  \begin{tabular}{ |c|c|c|c| }
    \hline
    Fiducial values & $\cal L$ [${\rm fb}^{-1}$] &  $B(\tilde{g}\to g\tilde{W}^0)$   & $B(\tilde{g}\to g\tilde{B}^0)$  \\ \hline
    $(B(tt\tilde{W}), B(bb\tilde{W}), B(tb\tilde{W}))  =(0.25, 0.25, 0.5)$ & 25 &0.01   &  0.02\\ \hline
    $B(qq\tilde{W})  =1$ & 25 &0.03   &  0.02\\ \hline  
    $B(qq\tilde{B})  =1 $ & 25 &0.02   &  0.06\\ \hline            
      \end{tabular}
      \end{center}
\end{table}

\section{Conclusion and Discussion}
{
The discovery of a 125 GeV Higgs boson suggests heavy scalar superpartner with mass ${\cal O} (10^{1-4})$ TeV. 
Direct search of such a heavy particle is impossible at the LHC. 
If the scalar partners are heavy enough,
an R-hadron or a displaced vertex are expected, 
which enable us to estimate the scale of the scalar sector.
}

{
However, if the scalar mass scale is much less than $ {\cal O} (10^{3})$ TeV, 
these features are not applicable.
Instead, we discussed that the gluino two body decay into a gluon 
is quite important to probe the scale of the scalar sector, since
the gluino two body decay is sensitive for the scalar mass scale.
The larger scalar mass scale results in enhancement the gluino two body decay.
Therefore the measurement of this branching fraction provides us information on the scalar mass scale,
which is an essential element for the Higgs mass estimation.
}

{
In this paper, we discuss measurability of this fraction at the LHC.
At parton level, there is an obvious difference between the two and three body decays.
In reality, it is quite non-trivial to discriminate the two body decay events.
We consider statistical discrimination.
The signal numbers after common-used event cuts,
strongly depends on the gluino decay modes, 
as seen Figs. \ref{fig:sig_simp}, \ref{fig:sig_split} and \ref{fig:sig_amsb}.
By using this feature, we show that measurement of the gluino branching fractions are possible.
For ${\cal O}(10^4)$ SUSY events, the branching fractions can be determined with accuracy of $\delta B \simeq 0.05$,
which corresponds to determination of the scalar scale with accuracy of a factor of 2.
Furthermore, the present method mainly relies on the event topology and not on the detailed of the mass spectrum.
Therefore this method does not depend on the details of the mass spectrum.
}

In order to make the measurement more accurate, more fine binning of the SUSY events can improve the discrimination by a few factor.
Or introducing more complicated variable such as $M_{\rm eff, 4}/M_{\rm eff, 2}$ distribution can also help.
Even though there is still room for improvement, the present expected accuracy of the branching fraction is good
enough to estimate the scalar mass scale.
Although our result is based on the fast simulation,
we expect that a full simulation can also reproduce the present result.

\section*{Acknowledgements}
The work of RS and KT is supported in part by JSPS Research Fellowships for Young Scientists.
This work is also supported by the World Premier International Research Center Initiative (WPI Initiative), MEXT, Japan.


\end{document}